\documentclass[aps,amsmath,amssymb,showpacs,lengthcheck]{revtex4}
\usepackage{epsfig}
\usepackage[english]{babel}
\usepackage{mhchem}
\usepackage{mathtools}

\begin{document}

\title{Comment on ``Nonlocal statistical field theory of dipolar particles in electrolyte solutions" by Y.A. Budkov}

\author{Sahin Buyukdagli$^{1}$\footnote{email:~\texttt{Buyukdagli@fen.bilkent.edu.tr}},  T. Ala-Nissila$^{2,3}$\footnote{email:~\texttt{Tapio.Ala-Nissila@aalto.fi}}, and Ralf Blossey$^{4}$\footnote{email:~\texttt{ralf.blossey@univ-lille1.fr}}}
\affiliation{$^{1}$Department of Physics, Bilkent University, Ankara 06800, Turkey\\
$^{2}$Department of Applied Physics and COMP Center of Excellence, Aalto University School of Science, 
P.O. Box 11000, FI-00076 Aalto, Espoo, Finland\\
$^{3}$Departments of Mathematical Sciences and Physics, Loughborough University, Loughborough, 
Leicestershire LE11 3TU, United Kingdom\\
$^{4}$University of Lille, Unit\'{e} de Glycobiologie Structurale et Fonctionnelle, CNRS UMR8576, 59000 Lille, France}

\maketitle

The article~\cite{budkov18} by Budkov introduces a nonlocal field-theoretic model of solvent-explicit electrostatics. Despite giving a detailed introduction to the early literature on the topic, the article misses out on a series of articles that we published several years ago. Consequently, Ref.~\cite{budkov18} essentially rederives without mention several results that were derived by us for the first time in Refs.~\cite{buyuk13-pre,buyuk13-jcp,buyuk14-jpcm,buyuk14-jcp}. The work by Budkov also considers variations of our model that are based on a different Yukawa-like solvent structure factor, which enables the exact evaluation of the average electrostatic potential.

In Refs.~\cite{buyuk13-pre,buyuk13-jcp,buyuk14-jpcm,buyuk14-jcp}, we developed the first field-theoretic model of nonlocal electrostatics embodying explicitly the extended charge structure of multipolar solvent molecules and ions with intrinsic polarizability. These articles showed that the consideration of the extended solvent charge structure allows to reproduce the non-local dielectric permittivity fluctuations observed in Molecular Dynamics simulations and solves the problem of the UV-divergence of the free energy without the introduction of an arbitrary cutoff.  

In Sec.2.1 of Ref.~\cite{budkov18}, the Author first derives the nonlocal field theoretic partition function from a formally more general expression by introducing a general solvent structure factor $g(r-r')$. Then, this generalization is dropped to switch to the dipolar case. The resulting field theoretic Hamiltonian Eq.(18) becomes exactly identical to the Hamiltonian functionals of our articles~\cite{buyuk14-jpcm} (Eq.(4)) and~\cite{buyuk14-jcp} (Eq.(1)), and the dipolar limit of the more general Hamiltonian of Ref.~\cite{buyuk13-pre} (Eq.(9)). The Author does not cite our previous works in any of these derivations.

The strong overlaps and similarities of Ref.~\cite{budkov18} with our published work extend beyond Sec.2.1. In Sec.2.2,  the dipole limit of the non-local Poisson-Boltzmann (NLPB) equation (22) corresponds to the MF-level NLPB equation of our article~\cite{buyuk14-jpcm} (Eq.(5)) and a restricted case of the more general NLPB equation in Ref.~\cite{buyuk13-pre} (Eq.(12)). Moreover, in Sec.2.3, the linearization of the NLPB equation and its solution in Fourier space (Eqs.(24)-(30))  bears strong similarities to our articles.  Most importantly, the dipolar dielectric function in Eq.(31) that follows from this solution is exactly identical to the dielectric permittivity function of Ref.~\cite{buyuk14-jpcm} (Eq.(10)). The latter also corresponds to the restricted case of the more general dielectric  function previously derived in Ref.~\cite{buyuk13-pre} (Eq.(21)). Again, no credit is given by the Author to our earlier articles. 

The overlaps of Ref.~\cite{budkov18} with our early works go beyond the MF treatment of the model. In Sec.2.5, the inverse Green's function in Eq.(56) is similar to Eq.(17) of our article~\cite{buyuk14-jcp}.  More precisely, the dipolar limit of Eq.(56) corresponds to the one-loop limit of the (variational) inverse kernel Eq.(17) of Ref.~\cite{buyuk14-jcp}. Finally, in the conclusion part of Ref.~\cite{budkov18}, the Author expresses his intention to generalize the theory by considering an "arbitrary electric structure", via the introduction of "a probability distribution function of distance for each pair of oppositely charged groups", without any citation to our non-local field theoretic models of Refs.~\cite{buyuk13-pre,buyuk13-jcp} where the ionic polarizability was taken into account within a Drude oscillator model.

In summary, we find it disconcerting that the most pertinent references to Ref.~\cite{budkov18} were not properly acknowledged. This is a severe shortcoming since the Author of Ref.~\cite{budkov18} knows about our work; he has already referred to our articles~\cite{buyuk13-pre,buyuk14-jcp} in his publications~\cite{budkov15,budkov16} on point-like dipoles.

\end{document}